\documentclass[final,5p,times]{elsarticle}

\usepackage{graphicx}

\usepackage{bm,amssymb,amsmath,amsfonts,xcolor}

\journal{Biochimica et Biophysica Acta -- Biomembranes}

\begin{document}

\begin{frontmatter}



\title{Membrane properties revealed by spatiotemporal response to a local inhomogeneity}


\author{Anne-Florence Bitbol\fnref{label2}} 
\author{Jean-Baptiste Fournier}

\address{Universit\'e Paris Diderot, Paris 7, Sorbonne Paris Cit\'e, Laboratoire Mati\`ere et Syst\`emes Complexes (MSC), UMR 7057 CNRS, B\^atiment Condorcet, Case Courrier 7056, F-75205 Paris Cedex 13, France}
\fntext[label2]{Current address: Lewis-Sigler Institute for Integrative Genomics, Princeton University, Princeton, New Jersey, United States of America}

\begin{abstract}
We study theoretically the spatiotemporal response of a lipid membrane submitted to a local chemical change of its environment, taking into account the time-dependent profile of the reagent concentration due to diffusion in the solution above the membrane. We show that the effect of the evolution of the reagent concentration profile becomes negligible after some time. It then becomes possible to extract interesting properties of the membrane response to the chemical modification. We find that a local density asymmetry between the two monolayers relaxes by spreading diffusively in the whole membrane. This behavior is driven by intermonolayer friction. Moreover, we show how the ratio of the spontaneous curvature change to the equilibrium density change induced by the chemical modification can be extracted from the dynamics of the local membrane deformation. Such information cannot be obtained by analyzing the equilibrium vesicle shapes that exist in different membrane environments in light of the area-difference elasticity model.
\end{abstract}


\end{frontmatter}



\section{Introduction}
During cell life, membranes are submitted to an inhomogeneous and variable environment. Local inhomogeneities can be strongly related to biological processes, which has led to experiments investigating the effect of local modifications on biomimetic membranes~\cite{Angelova99}. For instance, in the inner membrane of mitochondria, the enzymes that use the local pH difference across the membrane to synthesize adenosine triphosphate, the cell's fuel, are located in membrane invaginations called cristae~\cite{Davies11}. Experiments on model lipid membranes have shown that a local pH change can induce a local dynamical membrane deformation~\cite{Khalifat08,Fournier09,Bitbol11_guv,Khalifat11,Bitbol12}, and in particular the formation of cristae-like invaginations~\cite{Khalifat08}: membrane shape is tightly coupled to local pH inhomogeneities. Other concentration inhomogeneities in the environment of a cell have a crucial biological role, for instance in chemotaxis or in paracrine signaling. It is therefore of great interest to study the response of a biological membrane to a local modification of its environment.

Motivated by experiments conducted on biomimetic membranes, we have developed a theoretical description of the dynamics of a lipid bilayer membrane submitted to a local concentration increase of a substance that reacts reversibly and instantaneously with the membrane lipid headgroups. We focus on the regime of small deformations, and we treat linear membrane dynamics in the spirit of Ref.~\cite{Seifert93}. While our first works focused on the simple case of a constant modification of the membrane involving only one wavelength~\cite{Fournier09, Bitbol11_guv}, we have recently extended our theoretical description in order to take into account the spatiotemporal profile of the fraction of chemically modified lipids resulting from the local reagent concentration increase~\cite{Bitbol12}. This profile is determined by the diffusion of the reagent in the solution that surrounds the membrane. In Ref.~\cite{Bitbol12}, we compared the predictions of this theoretical description to experimental measurements of the deformation of the membrane of giant unilamellar vesicles caused by local microinjection of a basic solution, and we obtained good agreement between theory and experiment.

In the present article, we pursue the theoretical investigation of the effect of a local chemical modification on a lipid membrane. In general, the dynamics that results from a local reagent concentration increase is quite complex, as it involves the evolution of the reagent concentration profile simultaneously as the response of the membrane. This is the case in the experimental data analyzed in Ref.~\cite{Bitbol12}, which corresponds to microinjection steps lasting a few seconds. Here, we show that the effect of the evolution of the reagent concentration profile on the membrane dynamics becomes negligible some time after the beginning of the reagent concentration increase, after what the dynamics corresponds to the response of the membrane to a chemical modification imposed instantaneously. We find that studying this regime enables to extract interesting properties of the membrane response.

The article is organized as follows. First, in Sec.~\ref{Mbr_dyn}, we review the linear dynamics of a membrane submitted to a local chemical modification. Then, in Sec.~\ref{Res}, we study separately the dynamics associated with each of the two effects that can arise from a chemical modification, namely a spontaneous curvature change and an equilibrium density change of the external monolayer. We find that a local asymmetric density perturbation between the two monolayers of the membrane relaxes by spreading diffusively in the whole membrane. Intermonolayer friction plays a crucial part in this behavior. Subsequently, in Sec.~\ref{Res_b}, we treat the general case where both effects are present, and we show how the ratio of the spontaneous curvature change to the equilibrium density change induced by the local chemical modification can be extracted from the dynamics. This ratio cannot be deduced from the study of global modifications of vesicle equilibrium shapes in light of the area-difference elasticity model~\cite{Lee99}. Finally, Sec.~\ref{ccl} is a conclusion.

\section{Dynamics of a chemically modified membrane}
\label{Mbr_dyn}

For the article to be self-contained, the present section reviews the linear dynamics of a membrane submitted to a local chemical modification, starting from first principles. The main points of this description were presented in Ref.~\cite{Bitbol12}. In that article, we compared theoretical predictions to experimental measurements of the deformation of a membrane submitted to a local and brief pH increase. Here, our aim will be to go further in the analysis of our theoretical description in order to understand the fundamental properties of the response of a membrane to a local chemical modification.

\subsection{Monolayer free energy}
Our description of the bilayer membrane is based on a local version of the area-difference elasticity membrane model~\cite{Seifert93,Miao02,Bitbol11_stress}. 
We focus on small deformations of an infinite flat membrane, and we denote the upper monolayer by~$+$ and the lower one by~$-$. 

In the absence of a chemical modification, the local state of each monolayer is described by two variables: the local total curvature $c$ defined on the membrane midlayer, which is common to both monolayers, and the local scaled density $r^\pm=(\rho^\pm-\rho_0)/\rho_0$, defined on the midlayer of the membrane, $\rho_0$ being a reference density. The sign convention for the curvature is chosen in such a way that a spherical vesicle has $c<0$. The free energy $f^\pm$ per unit area in monolayer $\pm$ reads~\cite{Bitbol11_stress}:
\begin{equation}
f^\pm=\frac{\sigma_0}{2}+\frac{\kappa}{4}c^2\pm\frac{\kappa c_0}{2}c+\frac{k}{2}\left(r^\pm \pm ec\right)^2\,,
\label{fpm0}
\end{equation}
where $\sigma_0$ represents the tension of the bilayer and $\kappa$ its bending modulus, while $k$ is the stretching modulus of a monolayer, and $e$ denotes the distance between the neutral surface~\cite{Safran} of a monolayer and the midsurface of the bilayer. As we assume that the two monolayers of the membrane are identical before the chemical modification, these constants are the same for both monolayers. The spontaneous curvatures of the two monolayers have the same absolute value $c_0$ and opposite signs, since their lipids are oriented in opposite directions. The expression for $f^\pm$ in Eq.~(\ref{fpm0}) corresponds to a general second-order expansion in the small dimensionless local variables $r^\pm$ and $e c$, around the reference state which corresponds to a flat membrane with uniform density $\rho^\pm=\rho_0$. It is valid for small deformations around this reference state: $r^\pm=\mathcal{O}(\epsilon)$ and $ec=\mathcal{O}(\epsilon)$, where $\epsilon$ is a small dimensionless parameter used for bookkeeping purposes, which characterizes the amplitude of the small deformations of the membrane around the reference state. Mathematically, $\epsilon$ is considered infinitesimal. Note that in general, both $c$ and $r^\pm$, which describe local small deformations around the reference state, are functions of time and of position on the membrane.

Let us now focus on the way the membrane free energy is affected by the local chemical modification. We consider that the reagent source, which corresponds to the micropipette tip in an experiment, is localized in the water above the membrane. Besides, membrane permeation and flip-flop are neglected given their long timescales. Hence, the chemical modification only affects the upper monolayer, i.e., monolayer $+$, and not the lower one. Let us denote by $\phi$ the local mass fraction of the lipids of the upper monolayer that are chemically modified: $\phi$ depends on time and position since it arises from the local chemical modification. We assume that the reagent concentration is small  enough for $\phi$ to remain small at every time and position on the membrane, and we characterize this smallness through $\phi=\mathcal{O}(\epsilon)$.  In order to describe the chemically modified membrane, we have to include the third small variable $\phi$ in our second-order expansion of $f^+$. We obtain~\cite{Bitbol11_stress}: 
\begin{eqnarray}
f^+&=&\frac{\sigma_0}{2}+\sigma_1\phi+\frac{\sigma_2}{2}\phi^2+\tilde\sigma\left(1+r^+\right)\phi\ln\phi
+\frac{\kappa}{4}c^2\nonumber\\&&+\frac{\kappa}{2}\left(c_0+\tilde c_0\phi\right)c+\frac{k}{2}\left(r^++ ec\right)^2\,,
\label{fmod}
\end{eqnarray}
where the constants $\sigma_1$, $\sigma_2$, and $\tilde c_0$ describe the response of the membrane to the chemical modification. These constants depend on the reagent that is injected. Their physical meaning will be explained in the next paragraph. Besides, the non-analytical mixing entropy term $\tilde\sigma\left(1+r^+\right)\phi\ln\phi$ has been added to our second-order expansion~\cite{Bitbol11_stress}. Note that we assume that the three small dimensionless local variables $\phi$, $r^\pm$ and $ec$ are of the same order. In fact, in the present work, the deformation of the membrane and the density variation are caused by the local chemical modification, i.e., they are a response to $\phi$, which justifies that $ec$ and $r^\pm$ are of the same order as $\phi$. We refer the reader to Ref.~\cite{Bitbol11_stress} for more details on the derivation of Eqs.~(\ref{fpm0}) and~(\ref{fmod}). 

The effect of the chemical modification (i.e., of $\phi$) on the upper monolayer is twofold. First, the scaled equilibrium density on the neutral surface of the upper monolayer is changed by the amount $\sigma_1\phi/k$ to first order. Second, the spontaneous curvature of the upper monolayer is changed by the amount $-\bar c_0\phi$ to first order, with $\bar c_0=\tilde c_0+2\sigma_1 e/\kappa$. These results are obtained by minimization of the free energy of a homogeneous monolayer with constant mass (see~\ref{Ap_sc_dens}). Hence, the constants $\sigma_1$ and $\bar c_0$ describe the linear response of the monolayer equilibrium density and of its spontaneous curvature, respectively, to the chemical modification. This explains the physical meaning of the constants $\sigma_1$ and $\tilde c_0$ in Eq.~(\ref{fmod}). Note that $\sigma_2$ corresponds to the quadratic response of the membrane to the chemical modification, but it will not have any relevant effect in the following.

\subsection{Dynamical equations}

\begin{figure}[htb]
\centerline{\includegraphics[width=0.8\columnwidth]{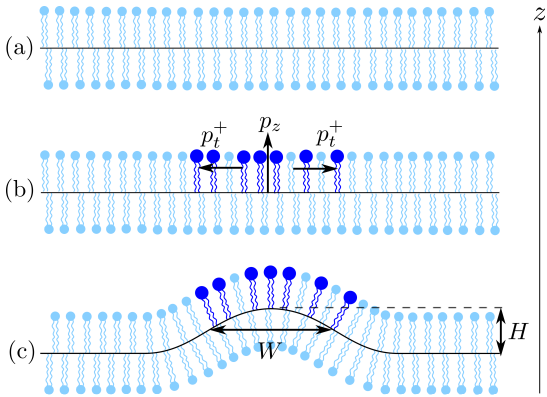}}
\caption{Qualitative representation of the early stages of the membrane response to an instantaneous chemical modification affecting the upper monolayer (monolayer $+$). (a): Membrane at equilibrium in the reference state (flat shape, uniform lipid densities in both monolayers). (b): Membrane just after the instantaneous local chemical modification: the membrane has not deformed yet at this stage, and the densities are still uniform. Some lipids of monolayer $+$ are modified: they are represented in dark blue ($\phi$ represents their local mass fraction; for clarity $\phi$ is locally large in the figure, but in our work, $\phi=\mathcal{O}(\epsilon)$). Here we assume that the result of the chemical modification is to increase the effective size of their headgroups. Hence, the equilibrium density in monolayer $+$ is locally decreased ($\sigma_1<0$), while the spontaneous curvature is increased in absolute value ($\bar c_0>0$). This results into a normal force density $p_z>0$ at the center of the modified zone of the membrane (see Eq.~(\ref{pn_b})), and into a tangential force density in monolayer $+$, oriented toward the exterior of the modified zone (see Eq.~(\ref{pip_b})). Both are indicated on the figure by arrows. (c) Due to the normal force density, the membrane starts deforming upwards. The maximum height $H$ and the full width at half maximum $W$ of the deformation are indicated. Note that tangential flows also appear due to tangential force densities.\label{Mem_fig}}
\end{figure}

The elastic force densities in a monolayer described by the free-energy densities in Eqs.~(\ref{fpm0}--\ref{fmod}) have been derived in Ref.~\cite{Bitbol11_stress} to first order in $\epsilon$, using the principle of virtual work. As we focus on small deformations of an infinite flat membrane, it is convenient to describe it in the Monge gauge by the height $z=h(\bm{r})$, $\bm{r}\in \mathbb{R}^2$, of its midlayer with respect to the reference plane $z=0$. Then, $ec=e\nabla^2 h$ to second order. Such a description is adapted to practical cases where the distance between the reagent source and the membrane is much smaller than the vesicle radius. The force per unit area of the reference plane, which we call ``force density'', then reads to first order in $\epsilon$
\begin{align}
\bm{p}_t^+&=-k\,\bm{\nabla}\left(r^++e\nabla^2 h-\frac{\sigma_1}{k}\phi\right)\,,\label{pip_b}\\
\bm{p}_t^-&=-k\,\bm{\nabla}\left(r^--e\nabla^2 h\right)\,,\label{pim_b}\\
p_z&=\sigma_0 \nabla^2 h-\tilde{\kappa}\nabla^4 h-k e\,\nabla^2 r_a-\left(\frac{\kappa \bar{c}_0}{2}-\sigma_1 e\right)\nabla^2\phi\,,
\label{pn_b}
\end{align}
where $\bm{p}_t^\pm$ is the tangential component of the force density in monolayer ``$\pm$", while $p_z=p_z^++p_z^-$ is the total normal force density in the membrane. In these formulas, we have introduced the antisymmetric scaled density $r_a=r^+-r^-$, and the constant $\tilde\kappa=\kappa+2ke^2$. These expressions show that both the equilibrium density change and the spontaneous curvature change (i.e., both $\sigma_1$ and $\bar c_0$) can yield a normal force density, and thus a deformation of the membrane, while only the equilibrium density change can yield a tangential force density and induce tangential lipid flow. An illustration of the role of these force densities in the membrane response to a local chemical modification is provided in Fig.~\ref{Mem_fig}.

Using the elastic force densities in Eqs.~(\ref{pip_b}--\ref{pn_b}), it is possible to describe the dynamics of the membrane to first order in the spirit of Ref.~\cite{Seifert93}. Let us outline the derivation of these equations~\cite{Bitbol11_guv, Bitbol12}, which is described in more detail in~\ref{Apeq}.
We use a normal force balance for the bilayer (see~\ref{Ap_mbdyn}), which involves $p_z$ and the normal viscous stress exerted by the surrounding fluid, which is derived in~\ref{Ap_hyd}. Besides, as each monolayer is a two-dimensional fluid, we write down generalized Stokes equations (see~\ref{Ap_mbdyn}), which involve $\bm{p}_t^\pm$, the two-dimensional viscous stress associated with the lipid flow, the tangential viscous stress exerted by the surrounding fluid (see~\ref{Ap_hyd}), and the intermonolayer friction $\mp b\left(\bm{v}^+-\bm{v}^-\right)$, where $\bm{v}^\pm$ is the velocity in monolayer $\pm$ and $b$ is the intermonolayer friction coefficient~\cite{Seifert93,Evans94}. This intermonolayer friction is a tangential force that comes into play when one monolayer slides relatively to the other one\footnote{Note that the two monolayers are supposed to remain in contact with one another at every time and point, due to hydrophobic forces: the sliding is only tangential. }. Finally, we use mass conservation in each monolayer. These dynamical equations are best expressed using two-dimensional Fourier transforms of the various fields involved, denoted with hats: for any field $f$ which depends on $\bm{r}$ and on time $t$, $\hat f$ is such that
\begin{equation}
\hat f(\bm{q},t)=\int_{\mathbb{R}^{2}}d\bm{r}\,f(\bm{r},t) e^{-i \bm{q}\cdot\bm{r}}\,.
\end{equation}
Combining all the above-mentioned equations yields a system of first-order linear differential equations on the two-dimensional variable $X=(q\,\hat h,\hat r_a)$: 
\begin{equation}
\frac{\partial X}{\partial t}(\bm{q},t)+M(q)\,X(\bm{q},t)=Y(\bm{q},t)\,,
\label{ED}
\end{equation}
where we have introduced the matrix which describes the dynamical response of the membrane~\cite{Seifert93, Bitbol11_guv}:
\begin{equation}
M(q)=\left(\begin{array}{cc}
\displaystyle\frac{\sigma_0 q+\tilde\kappa q^3}{4\eta}
&-\displaystyle\frac{keq^2}{4\eta}\\\\
-\displaystyle\frac{keq^3}{b}
&\displaystyle\frac{kq^2}{2b}
\end{array}\right).
\label{ED_defs}
\end{equation}
Here, we have assumed that $\eta_2q^2\ll b$ and $\eta q\ll b$. This is true for all the wavevectors with significant weight in $\hat\phi$, if the modified lipid mass fraction $\phi$ has a smooth profile with a characteristic width larger than 1 $\mu$m.
Indeed, $\eta=10^{-3}\,\mathrm{J\,s/m^3}$ for water, and typically $\eta_2=10^{-9}\,\mathrm{J\,s/m^2}$ and $b=10^9\,\mathrm{J\,s/m^4}$~\cite{Pott02,Shkulipa06}. Besides, the forcing term in Eq.~(\ref{ED}) reads~\cite{Bitbol11_guv}:
\begin{equation}
Y(\bm{q},t)=\left(\begin{array}{c}
\displaystyle\frac{\kappa \tilde c_0 q^2}{8\eta}\hat\phi(\bm{q},t)\\\\
\displaystyle\frac{\sigma_1 q^2}{2 b}\hat\phi(\bm{q},t)
\end{array}\right).
\label{ED_defs_2}
\end{equation}

Eqs.~(\ref{ED}-\ref{ED_defs}) show that the membrane deformation is coupled to the antisymmetric density: the symmetry breaking between the monolayers causes the deformation of the membrane. Here, the symmetry breaking is caused by the chemical modification of certain membrane lipids in the external monolayer, i.e., to the presence of $\phi$. And indeed, Eq.~(\ref{ED_defs_2}) shows that the forcing term in Eq.~(\ref{ED}) is proportional to $\hat\phi (\bm{q},t)$.

\subsection{Profile of the fraction of chemically modified lipids}
\label{Phi}

Before solving Eq.~(\ref{ED}), we need to determine $\hat \phi (\bm{q},t)$, which is involved in the forcing term $Y (\bm{q},t)$. The profile $\phi(\bm{r},t)$ of the mass fraction of chemically modified lipids in the external monolayer arises from the local reagent concentration increase. We focus on reagents that react reversibly with the membrane lipid headgroups. Besides, we assume that the reaction between the lipids and the reagent is diffusion-controlled (see, e.g., Ref.~\cite{Eigen64}). In other words, the molecular reaction timescales are very small compared to the diffusion timescales. For such a reversible diffusion-controlled chemical reaction, $\phi(\bm{r},t)$ is instantaneously determined by the local reagent concentration on the membrane, which results from the diffusion of the reagent in the fluid above the membrane. Note that these hypotheses are verified in the experiments analyzed in Ref.~\cite{Bitbol12}, where the reagent is sodium hydroxide.

We consider that the reagent source is localized in $(\bm{r}, z)=(\bm{0}, z_0>0)$, which would represent the position of the micropipette tip in an experiment (see Ref.~\cite{Bitbol12}). The cylindrical symmetry of the problem then implies that the fields involved in our description only depend on $r=|\bm{r}|$. We focus on the regime of small deformations $h(r,t)\ll z_0$, and we work at first order in $h(r,t)/z_0$. Besides, we study the linear regime where $\phi(r,t)$ is proportional to the reagent concentration on the membrane: denoting by $C(r,z,t)$ the reagent concentration field, we have $\phi(r,t)\propto C(r,h(r,t),t)$. To first order in the membrane deformation, this can be simplified into $\phi(r,t)\propto C(r,0,t)$. 

The field $C$ is determined by the diffusion of the reagent from the local source in the fluid above the membrane. In microinjection experiments, there is also a convective transport of the reagent due to the injection, but the P\'eclet number remains so small that diffusion dominates. Besides, since the membrane is a surface and $\phi\ll1$, we can neglect the number of reagent molecules that react with the membrane when calculating $C$. Hence, $C$ can be obtained by solving the diffusion equation 
\begin{equation}
\partial_t C-D\nabla^2 C=S\,, 
\end{equation}
where $D$ is the diffusion coefficient of the reagent in the fluid above the membrane, and the source term reads 
\begin{equation}
 S(\bm{r},z,t)=S_0\delta(\bm{r})\delta(z-z_0)\theta(t), 
\end{equation}
where $\theta$ is Heaviside's function. This corresponds to a constant injection flow from the source starting at $t=0$. In addition, the membrane imposes a Neumann boundary condition
$\partial_z C\left(r,h(r,t),t\right)=0$. This relation, which corresponds to a vanishing flux across the membrane, can be simplified to first order into 
\begin{equation}
\partial_z C\left(r,0,t\right)=0\,.
\end{equation}

The solution to this diffusion problem reads
\begin{align}
C\left(r,z,t\right)&=\int_0^{t} \!\!dt' \int_{\mathbb{R}^2} \!\!\!\!d\bm{r'} \int_0^{+\infty}\!\!\!\!\!\!\!\!dz'\,S\left(\bm{r'},z',t'\right)\times\nonumber\\
&\phantom{abcdefghijklmnopqr}G\left(|\bm{r}-\bm{r'}|,z,z',t-t'\right)\nonumber\\
&=S_0\int_0^{t} dt'\,G\left(r,z,z_0,t-t'\right)\,,
\label{cint}
\end{align}
where the causal Green function $G$ of our diffusion problem can be expressed using the method of images~\cite{Alastuey}:
\begin{equation}
G\left(r,z,z',t\right)=G_\infty \left(r,z-z',t\right)+G_\infty \left(r,z+z',t\right)\,,
\label{G1}
\end{equation}
where we have introduced the infinite-volume causal Green function of the diffusion equation
\begin{equation}
G_\infty \left(r,z,t\right)=\frac{\theta(t)}{\left(4\,\pi\,D\,t\right)^{3/2}}\,\exp\left(-\frac{r^2+z^2}{4\,D\,t}\right)\,.
\label{G2}
\end{equation}
Combining Eqs.~(\ref{cint}, \ref{G1}, \ref{G2}) provides an analytical expression for $C(r,z,t)$, and for $\hat C (q,z,t)$. Since $\hat\phi(q,t)\propto\hat C\left(q,0,t\right)$, we thus obtain an analytical expression for $\hat\phi$, which reads
\begin{equation}
\hat\phi\left(q,t\right)\propto\mathrm{erf}\left(q\sqrt{Dt}-\frac{z_0}{2\sqrt{Dt}}\right) \frac{\cosh\left(qz_0\right)}{qz_0}- \frac{\sinh\left(qz_0\right)}{qz_0}\,,
\label{phiq2}
\end{equation}
where erf denotes the error function. 

As $t\to\infty$, $\hat\phi(q,t)$ converges towards 
\begin{equation}
 \hat\phi_s(q)\propto \frac{e^{-qz_0}}{qz_0}\,. 
\label{phis}
\end{equation}  

\subsection{Resolution of the dynamical equations}
\label{Sec_Resol}
The membrane deformation $\hat h$ is given by the solution to Eq.~(\ref{ED}) with $\hat\phi$ expressed in Eq.~(\ref{phiq2}) and with the initial condition $X(q,t=0)=(0,0)$, corresponding to a non-perturbed membrane. Hence, we have
\begin{equation}
q\,\hat h(q,t)=\int_0^t ds\,\left[v_1e^{-\gamma_1t}A(s)+v_2e^{-\gamma_2t}B(s)\right].
\label{solution}
\end{equation} \\
Here, we have introduced the eigenvalues $\gamma_1$ and $\gamma_2$ of the matrix $M(q)$ (see Eq.~(\ref{ED_defs})), the associated eigenvectors $V_1=(v_1,w_1)$ and $V_2=(v_2,w_2)$, and the solutions $A$ and $B$ of the linear system 
\begin{equation}                                                                                                                                                                                  V_1A(t)e^{-\gamma_1t}+V_2B(t)e^{-\gamma_2t}=Y(q,t).                                                                                                                                                                                                                                   \end{equation}
The integral in Eq.~(\ref{solution}) and the inverse Fourier transform of $\hat h$ can be calculated numerically. This gives the spatiotemporal evolution of the membrane deformation $h$. 

\subsection{Dimensionless form of the description}
Both for physical understanding and for computational efficiency, it is interesting to put our description in dimensionless form. Using the Buckingham Pi theorem and choosing $z_0$ as the distance unit, $bz_0^2/k$ as the time unit and $\sigma_0 b^2 z_0^4/k^2$ as the mass unit, the ten parameters in our equations yield seven dimensionless numbers:
\begin{align}
L_1=\frac{1}{z_0}\sqrt{\frac{\kappa}{\sigma_0}}\,,\,\,L_2=\frac{e}{z_0}\,,\,\,\Sigma_0=
\frac{\sigma_0}{k}\,,\nonumber\\
\Delta=\frac{Db}{k}\,,\,\,\mu=\frac{\sigma_0bz_0}{\eta k}\,,\,\,\Sigma_1=\frac{\sigma_1}{k}\,,\,\,G=\bar
c_0z_0\,.
\label{nbadim}
\end{align}

The parameters $L_1$ and $L_2$ compare the characteristic lengthscales $\sqrt{\kappa/\sigma_0}$ and $e$ of the membrane to $z_0$, while $\Sigma_0$ is a dimensionless version of the membrane tension $\sigma_0$. Besides, the parameter $\Delta$ quantifies the importance of the reagent diffusion on the membrane dynamics. We will discuss briefly the effect of varying $\Delta$ in the following.

In the case of a lipid membrane in water, the only dimensional parameters that can span various orders of magnitudes are $\sigma_0$ and $z_0$. For $\sigma_0\geq10^{-8}\,\mathrm{N/m}$, i.e., for realistic membrane tensions~\cite{Evans90,Pecreaux04}, and for $z_0\geq 5\,\mu$m, we will see in the following that the only relevant dimensionless parameters in the dynamics of the membrane are $\mu$ and $\alpha=-\Sigma_0 L_1^2 G/(L_2 \Sigma_1)$. The first one, $\mu$, then corresponds to the ratio of the two eigenvalues of $M(q)$ for $q=1/(2z_0)$ (see Sec.~\ref{anin}): hence, it is a crucial element of the membrane dynamical response. The second one, $\alpha$, quantifies the relative weight of the spontaneous curvature change and of the equilibrium density change of the external monolayer due to the chemical modification (see Sec.~\ref{Res_b}). Note indeed that $\alpha\propto G/\Sigma_1$, and that $G$ and $\Sigma_1$ describe the two effects of the reagent on the membrane, through $\bar c_0$ and $\sigma_1$. The effect of varying $\mu$ and $\alpha$ will thus be discussed in the following.

For our numerical calculations, we initially take 
\begin{align}
L_1=10^{-1},\,L_2=10^{-4},\,\Sigma_0=10^{-6}, \nonumber\\
\Delta=21.25,\,\mu=10. 
\label{values}
\end{align}
These values correspond to the injection of NaOH in water ($D=2125\,\mu\mathrm{m^2/s}$~\cite{Cussler}) from a source at $z_0=10\,\mu$m above a floppy membrane with typical constants $\sigma_0=10^{-7}\,\mathrm{N/m}$ (see Ref.~\cite{Pecreaux04} and references therein), $\kappa=10^{-19}\,\mathrm{J}$, $k=0.1\,\mathrm{N/m}$ and $e=1$~nm~\cite{Safran}. The values of the intermonolayer friction coefficient $b$ and of the water viscosity $\eta$ are those given above. Note that the time unit $bz_0^2/k$ is then equal to one second. 

\section{Extreme cases}
\label{Res} 

In this Section, we will study the two extreme cases $G=0$ and $\Sigma_1=0$. In other words, we will study separately the dynamics associated with an equilibrium density change and with a spontaneous curvature change of the external monolayer. We will see that these two manifestations of the chemical modification of the membrane due to the reagent concentration increase yield different spatiotemporal evolutions of the membrane deformation.
The general case where both effects are present will then be discussed in Sec.~\ref{Res_b}.
                                                                                         
\emph{A priori}, the dynamics of the membrane deformation is quite complex, as it involves the evolution of the reagent concentration profile, due to diffusion, simultaneously as the response of the membrane.                                                                                                                                                                                                                                                                                                                                                                                                                                                                                                                                                                                                                    
In order to investigate the effect of the reagent diffusion on the dynamics of the deformation, and to see when the effect of the evolution of the reagent concentration profile becomes negligible, we will compare the two following cases: \\
(i) the realistic case where $\hat\phi(q,t)$ is given by Eq.~(\ref{phiq2}), \\
(ii) the theoretical case where the stationary modification $\phi_s$ in Eq.~(\ref{phis}) is imposed instantaneously: $\hat\phi(q,t)=\hat\phi_s(q)\theta(t)$. Note that this case corresponds to the limit of very large $\Delta$.

\subsection{Equilibrium density change only} 
\label{EqDensC}

Let us first focus on the case where $G=0$, in which only the equilibrium density of the upper monolayer is affected by the chemical modification. Fig.~\ref{HW_density} shows the evolution with dimensionless time $\tau=kt/(bz_0^2)$ of the membrane deformation height $H(\tau)=h(0,\tau)$ in front of the source, and of the full width at half-maximum $W(\tau)$ of the deformation (see Fig.~\ref{Mem_fig}(c) for an illustration of the definitions of $H$ and $W$). The cases (i) and (ii) introduced above are presented. On these graphs, we also show $\Phi(\tau)=\phi(0,\tau)/\phi_s(0)$ and the full width at half-maximum $W_\phi$ of $\phi$ in case (i). 

Fig.~\ref{HW_density} shows that the membrane undergoes a transient deformation that relaxes to zero while getting broader and broader. The local asymmetric density perturbation relaxes by spreading in the whole membrane. However, this process is slowed down by intermonolayer friction, so the density asymmetry is transiently solved by a deformation of the membrane~\cite{Sens04, Bitbol11_guv}. This deformation is downwards if $\Sigma_1>0$, and upwards if $\Sigma_1<0$. 

\subsubsection{Effect of reagent diffusion}

\begin{figure}[htb]
\centerline{\includegraphics[width=0.8\columnwidth]{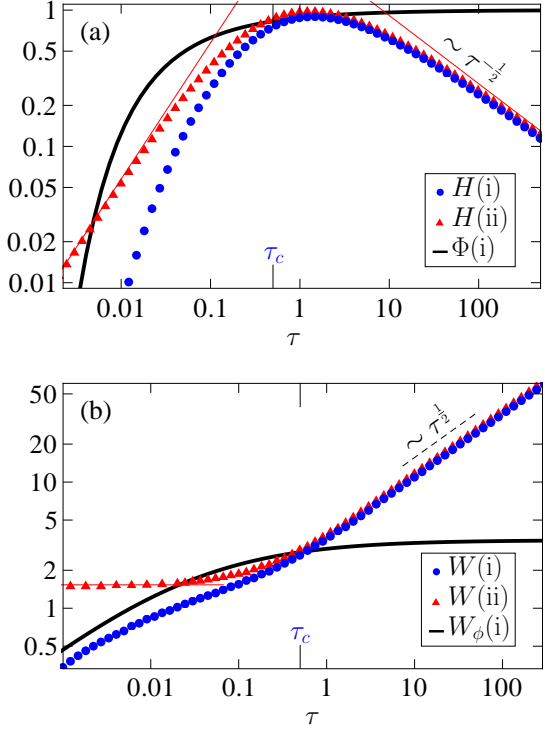}}
\caption{Dynamics of the membrane deformation in the extreme case $G=0$, where only the equilibrium density is changed. The values taken for the other dimensionless numbers are those in Eq.~(\ref{values}). Both the realistic case (i) where reagent diffusion is accounted for, and the simpler case (ii) where the chemical modification is imposed instantaneously, are considered. (a) Logarithmic plot of the height of the membrane deformation $H$ and of the fraction $\Phi$ of modified lipids in front of the reagent source versus dimensionless time $\tau$. Both in case (i) and in case (ii), $H$ is plotted in units of the extremal value it attains in case (ii). (b) Logarithmic plot of the width $W$ of the membrane deformation and of the width $W_\phi$ of the fraction of modified lipids versus $\tau$. Both $W$ and $W_\phi$ are plotted in units of $z_0$. It can be seen on graphs (a) and (b) that cases (i) and (ii) yield similar dynamics for $\tau\gg\tau_c\approx 0.5$. The thin red (gray) lines correspond to the analytical laws mentioned in the text. \label{HW_density}}
\end{figure}

In the realistic case (i), at short times, the membrane dynamics is governed by the evolution of $\phi$ due to the reagent diffusion. Conversely, at long times, once $\phi$ is close enough to its steady-state profile $\phi_s$, the dynamics of the membrane deformation is similar in case (i) and in case (ii). This can be seen in Fig.~\ref{HW_density}: first, $H$(i) and $W$(i) follow $\Phi$ and $W_\phi$, and then, they have an evolution very similar to those of $H$(ii) and $W$(ii). Thus, after some time, the effect of reagent diffusion on the dynamics of the membrane deformation becomes negligible, and the dynamics of the realistic case (i) can be well approximated by that of the simpler case (ii), which corresponds to the response of the membrane to an instantaneously imposed modification.

The transition time $\tau_c$ between the diffusion-dominated regime and the membrane-response--dominated regime is determined by the convergence of $\phi$ to $\phi_s$. As the reagent takes a dimensionless time $1/\Delta$ to diffuse from the source to the membrane, we expect $\tau_c\propto1/\Delta$. We studied the dynamics of the deformation height for $\Delta\in[10^{-3},10^3]$ by integrating Eq.~(\ref{ED}) numerically, and we found that this law is very well verified. For our standard value of $\Delta$ (see Eq.~\ref{values}), used in Fig.~\ref{HW_density}, we have $\tau_c\approx0.5\approx10/\Delta$.

\subsubsection{Analytical insight}
\label{anin}
Let us discuss analytically the simple case (ii). The long-time behaviors obtained in this case are especially interesting, since they also apply to the realistic case (i). Let us focus on $\sigma_0\geq10^{-8}\,\mathrm{N/m}$, i.e., on realistic membrane tensions, and let us keep the standard values of the other parameters involved in $M(q)$ (see below Eq.~(\ref{values})), as these parameters cannot vary significantly for a membrane in water. The eigenvalues of $M(q)$, which correspond to the two independent relaxation rates of a perturbation of the membrane with wavelength $2\pi/q$, can then be approximated by 
\begin{align}
\gamma_1\simeq\frac{kq^2}{2b}\,,\\
\gamma_2\simeq\frac{\sigma_0 q}{4\eta}\,,
\end{align}
for all wavevectors with significant weight in $\hat\phi_s$ if $z_0\geq 5\,\mu$m. Within this approximation, for $q=1/(2z_0)$, we have $\mu=\gamma_2/\gamma_1$. 
This leads to a simple expression of the solution of the dynamical equation Eq.~(\ref{ED}): 
\begin{equation}
 \hat  h (p,\tau)\propto\frac{e^{-p\mu\tau/4}-e^{-p^2\tau/2}}{\mu-2p}\,\hat\phi_s(p)\,,
\label{lim1}
\end{equation}
with $p=qz_0$. This expression shows that the only parameter that is relevant in the dynamics is $\mu$. Note that, in the realistic case (i), the value of $\Delta$ is relevant too. 

In the long-time limit, as the deformation spreads, we have $2p\ll\mu$ for all the wavevectors with significant weight in $\hat h(p,\tau)$. In this case, calculating the inverse Fourier transform of Eq.~(\ref{lim1}) for $r=0$ yields 
\begin{equation}
H\sim1/\sqrt{\tau}\,,
\end{equation}
and the numerical results for $W$ are in excellent agreement with 
\begin{equation}
 W\sim\sqrt{\tau}\,,
\end{equation}
as can be seen in Fig.~\ref{HW_density}(b). This law can also be obtained analytically if $\phi_s$ is replaced by a Gaussian. 

In the short-time limit, Eq.~(\ref{lim1}) yields $H\sim\tau$ and $W\to2(2^{2/3}-1)^{1/2}z_0$. These asymptotic behaviors, both for the short-time limit and for the much more interesting long-time limit, are plotted in red (gray) lines on Fig.~\ref{HW_density}. 

The transition between the two asymptotic regimes is determined by the value of $\mu$, which is the only parameter that controls the dynamics in case (ii). In particular, it is possible to show that the long-term scaling laws are valid for $\tau\gg\max\,(1,1/\mu^2)$. We studied the dynamics of the deformation height for $\mu\in[10^{-3},10^3]$ by integrating Eq.~(\ref{ED}) numerically, and we found that this law is very well verified. 

\subsubsection{Diffusive spreading of an antisymmetric density perturbation}

The long-term scaling $W\sim\sqrt{\tau}$ shows that the local antisymmetric density perturbation spreads diffusively. Let us emphasize that this diffusive behavior is not related to the diffusion of the reagent in the solution above the membrane, as in the long-term limit, $\phi$ has reached its steady-state profile $\phi_s$. The fact that the long-term scaling $W\sim\sqrt{\tau}$ holds in case (ii), where the profile $\phi_s$ is established instantaneously, as well as in case (i) (see Fig.~\ref{HW_density}(b)), illustrates that this scaling law is not related to the reagent diffusion. 

The long-term diffusive spreading of the antisymmetric density perturbation can be understood as follows. At long times, the difference between the Stokes equations for each monolayer Eq.~(\ref{baltg+}) and Eq.~(\ref{baltg-}) can be approximated by 
\begin{equation}
k\bm{\nabla}\left(r_a-\frac{\sigma_1}{k}\phi\right)+2b\left(\bm{v}^+-\bm{v}^-\right)=0\,,      
\label{balapp}                                                                                                                                                                                                                                                                      \end{equation}
in real space. To obtain this equation from the difference between Eq.~(\ref{baltg+}) and Eq.~(\ref{baltg-}), we have used $\eta_2q^2\ll b$ and $\eta q\ll b$ (see below Eq.~(\ref{ED_defs})), and also $|2e\,q^2\,\hat h|\ll |\hat r_a- \sigma_1\hat\phi/k|$, which holds for $\tau\gg1$. The last relation can be shown using Eq.~(\ref{lim1}) in the long-time limit. Using Eq.~(\ref{balapp}), the antisymmetric convective mass current $\bm{j}_a=\bm{j}^+-\bm{j}^-$ can be expressed to first order as 
\begin{equation}
\bm{j}_a=\rho_0\left(\bm{v}^+-\bm{v}^-\right)=-\rho_0\,\frac{k}{2b}\bm{\nabla}\left(r_a-\frac{\sigma_1}{k}\phi\right): 
\end{equation}
this current has a diffusive form. 
Combining this with the mass conservation relations in Eq.~(\ref{massc}) finally yields the diffusion equation 
\begin{equation}
 \partial_t r_a-\frac{k}{2b}\nabla^2\left(r_a-\frac{\sigma_1}{k}\phi\right)=0\,. 
\end{equation}
Hence, the local asymmetry in density between the two monolayers finally relaxes by spreading diffusively, with an effective diffusion coefficient $k/(2b)$. This slow relaxation is due to intermonolayer friction. Note that we can now interpret the dimensionless number $\Delta$ defined in Eq.~(\ref{nbadim}) as (half) the ratio of the reagent diffusion coefficient $D$ to this effective diffusion coefficient.

The fact that a local density asymmetry in the two monolayers of a membrane relaxes by spreading diffusively is generic. Here, the asymmetry is due to a local chemical modification of one monolayer, but it can also be caused, e.g., by a sudden flip of some lipids from one monolayer to the other~\cite{Sens04}. Our description of this diffusive behavior generalizes that of Ref.~\cite{Sens04}, which focused on a perturbation with a spherical cap shape and a uniform density asymmetry. 
Note that this diffusive behavior was first mentioned in Ref.~\cite{Evans94}, within a different theoretical framework, and restricting to axially-symmetric deformations.

\subsection{Spontaneous curvature change only} 
\label{SCC}
\begin{figure}[htb]
\centerline{\includegraphics[width=0.8\columnwidth]{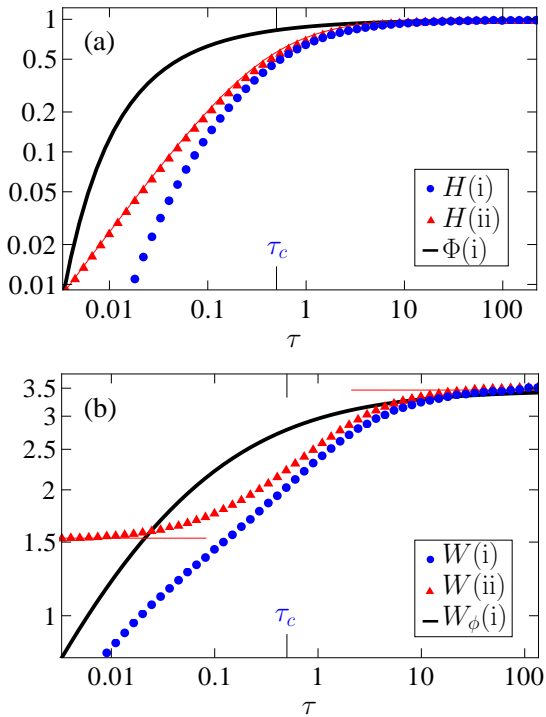}}
\caption{Dynamics of the membrane deformation in the extreme case $\Sigma_1=0$, where only the spontaneous curvature is changed. The values taken for the other dimensionless numbers are those in Eq.~(\ref{values}). Both the realistic case (i) where reagent diffusion is accounted for, and the simpler case (ii) where the chemical modification is imposed instantaneously, are considered. (a) Logarithmic plot of the height of the membrane deformation $H$ and of the fraction $\Phi$ of modified lipids in front of the reagent source versus dimensionless time $\tau$. Both in case (i) and in case (ii), $H$ is plotted in units of the extremal value it attains in case (ii). (b) Logarithmic plot of the width $W$ of the membrane deformation and of the width $W_\phi$ of the fraction of modified lipids versus $\tau$. Both $W$ and $W_\phi$ are plotted in units of $z_0$. It can be seen on graphs (a) and (b) that cases (i) and (ii) yield similar dynamics for $\tau\gg\tau_c\approx 0.5$. The thin red (gray) lines correspond to the analytical laws mentioned in the text. \label{HW_curv}}
\end{figure}

Let us now focus on the second extreme case, where $\Sigma_1=0$, i.e., where only the spontaneous curvature of the upper monolayer is affected by the chemical modification. Fig.~\ref{HW_curv} shows that the deformation converges to a deformed profile, in contrast with the previous case: while a local density asymmetry spreads on the whole membrane, a local spontaneous curvature modification leads to a locally curved equilibrium shape. The deformation is upwards if $G>0$, and downwards if $G<0$. 

As in Sec.~\ref{EqDensC}, in the realistic case (i), the membrane dynamics is governed by the reagent diffusion at short times $\tau\ll\tau_c$, while at long times $\tau\gg\tau_c$, it is similar in case (i) and in case (ii). The long-time dynamics thus corresponds to the response of the membrane to an instantaneously imposed chemical modification. 

In the simple case (ii), the approximations on $\gamma_1$ and $\gamma_2$ introduced in Sec.~\ref{anin} yield
\begin{equation}
\hat  h (p,\tau)\propto \left(1-e^{-p\mu\tau/4}\right)\hat\phi_s(p)\,,
\label{ancourb}
\end{equation}
where we have used the notation $p=qz_0$ as above. Calculating the inverse Fourier transform of this function for $r=0$ yields 
\begin{equation}
 H\propto \frac{\mu\tau}{\mu\tau+4}\,,
\label{anacourb_b}
\end{equation}
for all $\tau>0$. This analytical law for the deformation height $H$ is plotted in red (gray) lines in Fig.~\ref{HW_curv}(a). 

Let us focus on the long-time limit, which is especially interesting, since the results found in case (ii) also apply to the realistic case (i). Eq.~(\ref{ancourb}) shows that in this limit, $\hat  h (p,\tau)\propto\hat\phi_s(p)$, so that the long-time profile of the membrane deformation is fully determined by that of $\phi$. In particular, for $\tau\to\infty$, $W_\phi\to 2\sqrt{3}\,z_0$ and $W\to 2\sqrt{3}\,z_0$. 

For $\tau\to 0$, we find $W\to2(2^{2/3}-1)^{1/2}z_0$ again. These asymptotic behaviors regarding the deformation width $W$ are plotted in red (gray) lines in Fig.~\ref{HW_curv}(b), both for the short-time limit and for the much more interesting long-time limit. 

As in Sec.~\ref{EqDensC}, in case (ii), $\mu$ is the only relevant parameter in the dynamics of the membrane deformation (see Eq.~(\ref{ancourb})), and it determines the transition time between the two asymptotic regimes. In the realistic case (i), the value of $\Delta$ is relevant too. 

\section{General case} 
\label{Res_b}
Generically, a chemical modification will affect both the equilibrium density and the spontaneous curvature of the upper monolayer. The general solution of Eq.~(\ref{ED}) is a linear combination of the two solutions obtained for the two extreme cases: $G=0$ (see Sec.~\ref{EqDensC}) and $\Sigma_1=0$ (see Sec.~\ref{SCC}). In this Section, we will only discuss the realistic case (i) where the reagent diffusion is taken into account. As in the two extreme cases, the effect on the membrane dynamics of the evolution of $\phi$ due to reagent diffusion is crucial for $\tau\ll\tau_c$, while it becomes negligible for $\tau\gg\tau_c$.

For a lipid membrane in water such that $\sigma_0\geq10^{-8}\,\mathrm{N/m}$ and for $z_0\geq5\,\mu$m, we have seen that in both extreme cases, the dynamics is influenced only by $\mu$ and $\Delta$ (see Secs.~\ref{EqDensC}--\ref{SCC}). Hence, in the general case, the dynamics is influenced by $\mu$, $\Delta$, and by the parameter
\begin{equation}
\alpha=-\frac{\kappa\,\bar c_0}{\sigma_1 e}=-\frac{\Sigma_0 L_1^2}{L_2}\,\frac{G}{\Sigma_1}\,,
\end{equation}
which quantifies the relative importance of the spontaneous curvature change and of the equilibrium density change of the external monolayer due to the chemical modification. Indeed, Eq.~(\ref{pn_b}) shows that the ratio of the normal force density arising from the spontaneous curvature change to the normal force density arising from the equilibrium density change is equal to $-2\alpha$. The effect of varying $\Delta$ was discussed in Secs.~\ref{EqDensC}--\ref{SCC}, and in addition, $\Delta$ cannot vary much for a reagent injected in water above a lipid membrane. Hence, we will focus on the influence of $\mu$ and $\alpha$ on the membrane dynamics.

The value of the ratio $\alpha$ of the spontaneous curvature change to the equilibrium density change is \emph{a priori} unknown, as $\sigma_1$ and $\bar c_0$, which are involved in $\Sigma_1$ and $G$, respectively (see Eq.~\ref{nbadim}), are unknown. The actual values of $\sigma_1$ and $\bar c_0$ depend on the reagent as well as on the membrane itself, as these two parameters describe the linear response of the membrane to a reagent (see Eq.~\ref{fmod}). Let us assume that $\alpha>0$, i.e., that the equilibrium density change and the spontaneous curvature change induce deformations in the same direction. This is true, e.g., for a chemical modification that affects the lipid headgroups in such a way that it yields an effective change of the preferred area per headgroup~\cite{Bitbol11_guv}. A rough microscopic model, where lipids are modeled as cones that favor close-packing, then yields $\alpha\approx1$, which corresponds to the case where the two effects yield destabilizing normal force densities of similar magnitudes~\cite{Bitbol11_guv}. Thus, we expect $\alpha\approx 1$. 

\begin{figure}[htb]
\centerline{\includegraphics[width=\columnwidth]{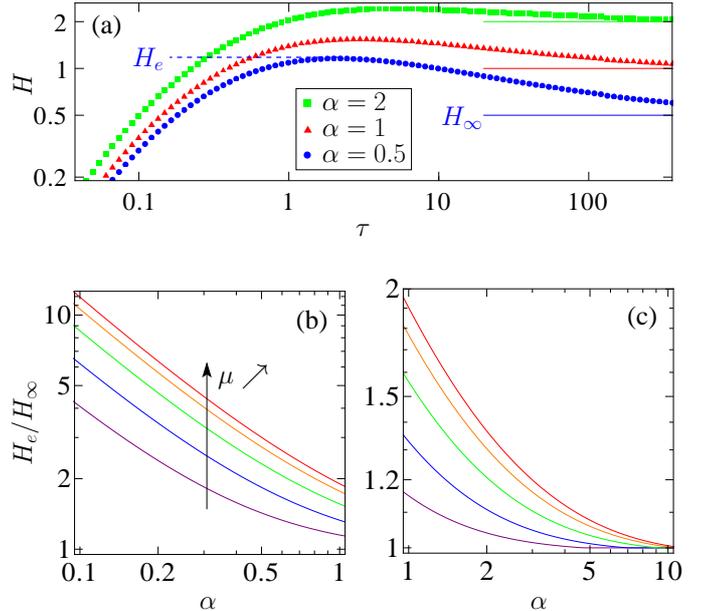}}
\caption{(a): Logarithmic plot of the height of the membrane deformation $H$ in front of the reagent source versus dimensionless time $\tau$ in the realistic case (i), for different values of the parameter $\alpha$, which quantifies the relative importance of the spontaneous curvature change to the equilibrium density change. The values taken for the other dimensionless numbers are those in Eq.~(\ref{values}). In each case, $H$ is plotted in units of its asymptotic value in the case $\alpha=1$. The lines show the asymptotic values $H_\infty$ of $H$ for $\tau\to\infty$ in each case, and $H_e$ denotes the extremal value of $H$. (b) and (c): Logarithmic plot of $H_e/H_\infty$ versus $\alpha$ in case (i), for different values of the dimensionless parameter $\mu$, which affects the membrane dynamics as well as $\alpha$. From down to up: $\mu=1;\,\,\mu=10^{0.5}\simeq 3.2;\,\,\mu=10;\,\,\mu=10^{1.5}\simeq 32;\,\,\mu=10^2$.
\label{hw_mixte}}
\end{figure}

Fig.~\ref{hw_mixte}(a) shows the evolution of $H$ in case (i) for three different values of $\alpha$. The deformation height features an extremum $H_e$, and then a relaxation. This behavior is due to the change of the equilibrium density (see Sec.~\ref{EqDensC}). Conversely, the nonzero asymptotic deformation $H_\infty$ arises from the change of the spontaneous curvature (see Sec.~\ref{SCC}). The relative importance of $H_e$ to $H_\infty$ thus depends on $\alpha$. This means that studying the dynamics of the membrane deformation in response to a local chemical modification provides information on the ratio of the spontaneous curvature change to the equilibrium density change induced by this chemical modification. 

In Fig.~\ref{hw_mixte}(b)--(c), the ratio $H_e/H_\infty$ is plotted versus $\alpha$ for different values of $\mu$. Indeed, as mentioned above, the membrane dynamics is determined both by $\alpha$ and by $\mu$ (at constant $\Delta$). The order of magnitude of $H_e$ can be estimated assuming an equilibrium density change and/or a spontaneous curvature change of a few percent: $H_e$ is of order $1-10\,\mu$m for flaccid membranes, and smaller than $0.1\,\mu$m if $\sigma_0\geq10^{-5}\,\mathrm{N/m}$. Hence, we choose $\sigma_0\in[10^{-8},10^{-6}]\,\mathrm{N/m}$, which yields $\mu\in[1,100]$ for $z_0=10\,\mu$m: such values are taken in Fig.~\ref{hw_mixte}(b)--(c). 

Our study shows that one can deduce $\alpha$, and thus the relative importance of the spontaneous curvature change to the equilibrium density change due to a chemical modification, from the measurement of $H_e/H_\infty$. This is very interesting, given that such information cannot be deduced from the study of static and global membrane modifications~\cite{Lee99,Miao94}. Indeed, the equilibrium vesicle shapes in the area-difference elasticity model are determined by the combined quantity
\begin{equation}
\overline{\Delta a_0}=\Delta a_0+\frac{2}{\xi}c_0^b\,,
\label{combqty}
\end{equation}
which involves both the equilibrium density and the spontaneous curvature. In this expression, $\Delta a_0$ is the dimensionless preferred area difference between the two monolayers, which is related to the asymmetric equilibrium density change, while $c_0^b$ denotes the dimensionless spontaneous curvature of the bilayer, which is related to the asymmetric spontaneous curvature change~\cite{Miao94}. The parameter $\xi$ in Eq.~(\ref{combqty}) is a dimensionless number involving the elastic constants of the membrane. Note that the vesicle shape variations due to global modifications of the vesicle environment are usually interpreted as coming only from a change of the spontaneous curvature, under the assumption that the preferred area per lipid is not modified~\cite{Lee99,Dobereiner99,Petrov99}.

In order to determine the ratio $\alpha$ of the spontaneous curvature change to the equilibrium density change  in a practical case, it is necessary to know the value of $\mu$ (see Fig.~\ref{hw_mixte}(b)--(c)). However, as $\mu$ does not involve any parameter that depends on the reagent (see Eq.~(\ref{nbadim})), it is possible to compare the effects of different reagents on the same membrane, i.e., their values of $\alpha$, even without knowing the precise value of $\mu$.

\section{Conclusion}
\label{ccl}
We have analyzed theoretically the spatiotemporal response of a membrane submitted to a local heterogeneity, in the realistic case of a local concentration increase of a substance that reacts reversibly and instantaneously with the membrane lipid headgroups. In general, the dynamics of the membrane deformation is quite complex, as it involves the evolution of the reagent concentration profile due to diffusion in the solution above the membrane, simultaneously as the response of the membrane. We have shown that, some time after the beginning of the reagent concentration increase, the effect of the evolution of the reagent concentration becomes negligible. 

Studying this regime enables to extract interesting properties of the membrane response. We have shown that a local density asymmetry between the two monolayers relaxes by spreading diffusively in the whole membrane. Intermonolayer friction plays a crucial part in this behavior. In addition, we have shown how the relative importance of the spontaneous curvature change to the equilibrium density change can be extracted from the dynamics of the membrane response to the local chemical modification. This is a significant result since such information cannot be deduced from the study of a static and global modification using the area-difference elasticity model.

Our description provides a theoretical framework for experiments involving the microinjection of a reagent close to biomimetic membranes. In Ref.~\cite{Bitbol12}, we used the theoretical model presented here to analyze experimental results corresponding to brief microinjections of a basic solution in the regime of small deformations, and we obtained good agreement between theory and experiment.
The results of the present theoretical work show that it would be interesting to conduct experiments with a continuous injection phase, since the ratio $\alpha$ of the spontaneous curvature change to the equilibrium density change could then be determined for various membrane compositions and reagents.

In biomimetic membranes as well as in cells, remarkable phenomena occur in the regime of larger deformations: cristae-like invaginations~\cite{Khalifat08}, tubulation~\cite{Fournier09}, pearling~\cite{Tsafrir03}, budding, exo- or endocytosis~\cite{Angelova99}, etc. To study such phenomena, it would be useful to pursue our study in the nonlinear regime.

\section*{Acknowledgements}
We thank Miglena I. Angelova for motivating us to study this subject. We thank Miglena I. Angelova and Nicolas Puff for interesting discussions.

\appendix
\section{Spontaneous curvature and equilibrium density}
\label{Ap_sc_dens}
The spontaneous curvature and the equilibrium density in monolayer $+$ can be obtained by minimizing $f^+/\rho^+$ with respect to $r^+$ and $c$ for a homogeneous monolayer with constant mass. First, using the expression of $f^+$ in Eq.~(\ref{fmod}), the minimization with respect to $r^+$ gives, to first order in $\epsilon$:
\begin{equation}
r^+_\mathrm{eq}=\frac{\sigma_0}{2\,k}+\frac{\sigma_1\phi}{k}-ec\,.
\label{min1}
\end{equation}
Then, the minimization with respect to $c$ yields to first order, using Eq.~(\ref{min1}):
\begin{equation}
c_\mathrm{eq}=-c_0-\bar c_0 \phi -\frac{\sigma_0 e}{\kappa}\,,
\label{min2}
\end{equation}
with $\bar c_0=\tilde c_0+2\sigma_1 e/\kappa$ (as defined in the main text). Note that, since we assume that $r^+=\mathcal{O}(\epsilon)$ and $ec=\mathcal{O}(\epsilon)$, we must have $c_0e=\mathcal{O}(\epsilon)$ and $\sigma_0/k=\mathcal{O}(\epsilon)$ for our description to be valid for the values of $r^+$ and $c$ that minimize $f^+/\rho^+$. This property has been used to simplify the results of the minimization.

The scaled lipid density $r_n^+$ on the neutral surface of the monolayer is related to $r^+$ through $r^+_n=r^++ec$ to first order. This relation arises from the geometry of parallel surfaces~\cite{Safran}, given that the membrane midlayer and the monolayer neutral surface are parallel surfaces separated by a distance $e$. Hence, Eq.~(\ref{min1}) can be rewritten as
\begin{equation}
r_{n,\,\mathrm{eq}}^+=\frac{\sigma_0}{2\,k}+\frac{\sigma_1\phi}{k}\,.
\label{min1b}
\end{equation}
This result is independent of the curvature $c$, contrary to that in Eq.~(\ref{min1}). Indeed, by definition, on the neutral surface, curvature and density are decoupled~\cite{Safran}, while these two variables are coupled on other surfaces. 

Eq.~(\ref{min1b}) shows that, due to the chemical modification, the scaled equilibrium density on the neutral surface of monolayer $+$ is changed by the amount 
\begin{equation}
 \delta r^+_{n,\,\mathrm{eq}}=r^+_{n,\,\mathrm{eq}}(\phi)-r^+_{n,\,\mathrm{eq}}(0)=\frac{\sigma_1\phi}{k}
\end{equation}
to first order. Besides, Eq.~(\ref{min2}) indicates that the spontaneous curvature of monolayer $+$ is changed by the amount
\begin{equation}
 \delta c_{\mathrm{eq}}=c_{\mathrm{eq}}(\phi)-c_{\mathrm{eq}}(0)=-\bar c_0\phi
\end{equation}
to first order.

One might wonder why the spontaneous curvature $c_\mathrm{eq}$ found by minimization is not simply $-c_0$ for $\phi=0$ (see Eq.~(\ref{min2})). This is due to the fact that we work on the membrane midsurface, which is more convenient to study the membrane dynamics. If the monolayer free energy had been originally written using variables and constants defined on the neutral surface, then the spontaneous curvature found by minimization would correspond exactly to the constant $c_0^n$ which plays the part of $c_0$ when everything is defined on the neutral surface.

\section{Derivation of the dynamical equations}
\label{Apeq}

\subsection{Hydrodynamics of the fluid above and below the membrane}
\label{Ap_hyd}

We wish to determine the velocity field $\bm{W}^\pm=(\bm{w}^\pm,W_z^\pm)$ in the fluid above ($+$) and below ($-$) the membrane. This flow is caused by the deformation of the membrane and by the lateral flow in the membrane: mathematically, it is determined by the boundary conditions corresponding to the continuity of velocity at the interface between the fluid and the membrane. 

Given the short lengthscales considered, the dynamics of $\bm{W}^\pm(\bm{r},z,t)$ can be described using Stokes' equation. Adding the incompressibility condition, we have:
\begin{align}
\eta (-q^2+\partial_z^2)\bm{\hat W^\pm}=(i\bm{q}+\bm{e}_z \partial_z)\hat P^\pm\,,\label{stokes}\\
i\bm{q}\cdot \bm{\hat w}^\pm+\partial_z \hat W_z^\pm=0\,,\label{inc}
\end{align}
where $P^\pm$ is the excess pressure field in the fluid: the total pressure is $\mathcal{P}^\pm=P_0+P^\pm$, where $P_0$ is a constant and $P^\pm$ goes to zero far from the membrane. As in the main text, hats indicate two-dimensional Fourier transforms. Taking the scalar product of $\bm{q}$ and Eq.~(\ref{stokes}), and using Eq.~(\ref{inc}) yields 
\begin{equation}
\hat P^\pm=-\eta \partial_z \hat W_z^\pm +\frac{\eta}{q^2}\partial_z^3 \hat W_z^\pm\,.
\label{interm}
\end{equation}
Taking the scalar product of $\bm{e}_z$ and Eq.~(\ref{stokes}), and using Eq.~(\ref{interm}) yields 
\begin{equation}
q^4 \hat W_z^\pm-2\,q^2 \partial_z^2 \hat W_z^\pm+\partial^4_z \hat W_z^\pm=0\,.
\label{edwz}
\end{equation}
Solving these equations, with the boundary conditions at infinity: $\hat W_z^\pm\to0$ for $z\to\pm\infty$, and at the water--membrane interface: $\hat W_z^\pm(\bm{q},0,t)=\partial_t \hat h(\bm{q},t)$ and $\bm{\hat w}^\pm(\bm{q},0,t)=\bm{\hat v}^\pm(\bm{q},t)$, finally yields
\begin{align}
\hat P^\pm&=\pm2\,\eta\left[q \,(\partial_t \hat h) \mp i \bm{q}\cdot\bm{\hat v}^\pm\right]\,e^{\mp q z}\,,\label{3d_1}\\
\bm{\hat w}^\pm&=\left\{\bm{\hat v}^\pm-\left[i\bm{q} \,(\partial_t \hat h) \pm q\, \bm{\hat v}^\pm\right] z\,\right\}e^{\mp q z}\,,\label{3d_2}\\
\hat W_z^\pm&=\left\{\partial_t \hat h\pm\left[q \,(\partial_t \hat h) \mp i \bm{q}\cdot\bm{\hat v}^\pm\right] z\,\right\}e^{\mp q z}\,.\label{3d_3}
\end{align}

For our study of the dynamics of the membrane, what is needed is the stress exerted by the fluid on the membrane, i.e., since we are working at first order in the membrane deformation, the stress in $z=0$. The viscous stress tensor of the fluid is defined by
\begin{equation}
T_{\alpha\beta}=-(P^\pm+P_0)\delta_{\alpha\beta}+\eta\left(\partial_\beta W^\pm_\alpha+\partial_\alpha W^\pm_\beta\right)\,,
\end{equation}
where $\alpha\in\{x,y,z\}$ and $\beta\in\{x,y,z\}$. Hence, using Eqs.~(\ref{3d_1}, \ref{3d_2}, \ref{3d_3}), we obtain
\begin{align}
\hat T^+_{zz}(z=0)-\hat T^-_{zz}(z=0)&=-4\,\eta\,q\,\partial_t \hat h\,,\label{st1}\\
\bm{\hat T}_{tz}^\pm(z=0)&=\mp 2\,\eta\,q\,\bm{\hat v}^\pm\,,\label{st2}
\end{align}
where we have introduced the tangential part $\bm{T}_{tz}^\pm=(T_{xz}^\pm,T_{yz}^\pm)$ of the stress tensor of the fluid above and below the membrane.

\subsection{Membrane linear dynamics}
\label{Ap_mbdyn}

The first dynamical equation we use is a balance of forces per unit area acting normally to the membrane. It involves the normal elastic force density in the membrane given by Eq.~(\ref{pn_b}) and the normal viscous stresses exerted by the fluid above and below the membrane, which are derived in~\ref{Ap_hyd} (see Eq.~(\ref{st1})). It reads:
\begin{equation}
-\left(\sigma_0 \,q^2+\tilde\kappa\, q^4\right)\hat h +k \,e\, q^2\,\hat r_a + \frac{\kappa \,\tilde{c}_0}{2}\,q^2\hat \phi -4\,\eta \,q\,\partial_t \hat h=0\,,
\label{balnorm}
\end{equation}
where $\eta$ denotes the viscosity of the fluid above and below the membrane.

Besides, as each monolayer is a two-dimensional fluid, we write down generalized Stokes equations describing the tangential force balance in each monolayer. The first force involved is the density of elastic forces given by Eqs.~(\ref{pip_b}) and (\ref{pim_b}). The second one arises from the viscous stress in the two-dimensional flow of lipids. The third one comes from the viscous stress exerted by the water, which is derived in~\ref{Ap_hyd} (see Eq.~(\ref{st2})). The last force that has to be included is the intermonolayer friction~\cite{Evans94}. We thus obtain:
\begin{align}
 -i\,k\,\bm{q}\left(\hat r^+ - e\,q^2\,\hat h - \frac{\sigma_1}{k}\,\hat\phi\right) &-\left(\eta_2\,q^2 +2\,\eta\,q\right)\bm{\hat v}^+ \nonumber\\
&-b\left(\bm{\hat v}^+ - \bm{\hat v}^-\right)=0\,,\label{baltg+}\\
 -i\,k\,\bm{q}\left(\hat r^- + e\,q^2\,\hat h\right) &-\left(\eta_2\,q^2 +2\,\eta \,q\right)\bm{\hat v}^- \nonumber\\
& +b\left(\bm{\hat v}^+ - \bm{\hat v}^-\right)=0\,,\label{baltg-}
\end{align}
where $\bm{v}^\pm$ denotes the velocity in monolayer $\pm$, while $\eta_2$ is the two-dimensional viscosity of the lipids and $b$ is the intermonolayer friction coefficient~\cite{Evans94}.

Finally, we use the conservation of mass in each monolayer to first order:
\begin{equation}
\partial_t \hat r^\pm + i\,\bm{q}\cdot\bm{\hat v}^\pm=0\,. \label{massc}
\end{equation}

Combining Eqs.~(\ref{balnorm}-\ref{massc}), we obtain Eqs.~(\ref{ED}-\ref{ED_defs_2}).





\bibliographystyle{model1-num-names}







\end{document}